\documentstyle[epsfig]{article}
\def\bc{\begin{center}}
\def\ec{\end{center}}
\def\beq{\begin{equation}}
\def\eeq{\end{equation}}

\def\sig {\sigma}

\def\hs#1{\hspace*{#1cm}}

\def\av#1#2{\langle {#1} \rangle^{}_{#2}}
\def\ave#1{\langle {#1} \rangle}

\def\F{{n^{}_F}}
\def\B{{n^{}_B}}
\def\nF{{n^{}_F}}
\def\nB{{n^{}_B}}
\def\pF{{p^{}_{{\rm t}F}}}
\def\pB{{p^{}_{{\rm t}B}}}
\def\ppF{{p^{2}_{{\rm t}F}}}
\def\ppB{{p^{2}_{{\rm t}B}}}
\def\ppiF{{p^{2}_{{\rm t}iF}}}
\def\ppiB{{p^{2}_{{\rm t}iB}}}

\def\DF{\Delta y_F^{}}
\def\DB{\Delta y_B^{}}

\def\bfs{{\bf s}}
\def\bfb{{\bf b}}

\def\j#1#2#3#4{{#1} {\bf #2} (#3) #4}

\def\NP{Nucl. Phys.}
\def\PL{Phys. Lett.}
\def\PRL{Phys. Rev. Lett.}
\def\PR{Phys. Rev.}

\def\PRep{Phys. Rep.}

\def\EPJ{Eur. Phys. J.}

\begin{document}

\begin{center}
{\Large \bf Multiplicity and $p_t$ correlations in AA-interactions\\ at high energies} \\

\vspace{4mm}

R.S. Kolevatov and V.V. Vechernin\\
V.A.Fock Institute of Physics, St.Petersburg State University\\
198504 St.Petersburg, Ulyanovskaya 1 \\
\end{center}

\begin{abstract}
The theoretical description of the correlations between observables
in two separated rapidity intervals in relativistic nuclear collisions
is presented.
It is shown, that the event-by-event $p_t$--$p_t$ correlation defined as the correlation
between event mean values of transverse momenta of all particles emitted
in two different rapidity intervals does not decrease to zero
with the increase of the number of strings
in contrast with two particle $p_t$--$p_t$ correlation -
the correlation between the transverse momenta
of single particles produced in these two rapidity windows.

In the idealized case with the homogeneous string distribution in the transverse plane
in the framework of the cellular analog of string fusion model (SFM)
the asymptotic of $p_t$-$p_t$ correlation coefficient
is analytically calculated and compared with
the results of the Monte-Carlo (MC) calculations
fulfilled both in the framework of the original SFM and in the framework of its cellular analog,
which enables to control the MC algorithms.

In the case with the realistic nucleon distribution density of colliding nuclei
the results of the MC calculations of the $p_t$--$p_t$ correlation function
for minimum bias nuclear collisions at SPS, RHIC and LHC energies
are presented and analysed.
\end{abstract}

\section{String fusion model (SFM)}

The colour string model \cite{Capella1,Kaidalov}
originating from Gribov-Regge approach
is being widely applied
for the description of the soft part of the multiparticle production
in hadronic and nuclear interactions at high energies.
In this model
at first stage of hadronic interaction
the formation of the extended objects -
the quark-gluon strings - takes place.
At second stage
the ha\-d\-ro\-ni\-za\-tion of these strings produces the observed hadrons.
In the original
version the strings evolve independently and the observed spectra are just
the sum of individual string spectra.
However in the case of nuclear collision,
with growing energy and atomic number of colliding nuclei,
the number of strings grows and
one has to take into account the interaction between them.

One of possible approaches to the problem
is the colour string fusion model \cite{BP1}.
The model is based on a simple observation
that due to final transverse dimensions of strings
they inevitably have to start to overlap
with the rise of their density in transverse plane.
At that the interaction of string colour fields takes place,
which changes the process of their fragmentation into hadrons
as compared with the fragmentation of independent strings.
So we have one more interesting nonlinear phenomenon
in nuclear interactions at high energies - the field of physics
the investigations in which were initiated by pioneer works
of academician A.M.~Baldin \cite{Baldin}.

It was shown \cite{BP1,BP00EPJC,BP00PRL} that the string fusion phenomenon considerably damps
the charged particle multiplicity and simultaneously
increase their mean $p_t$ value
as compared with the case of independent strings.
In accordance with a general Schwinger idea \cite{Schwinger51}
and the following papers \cite{Biro,Bialas} (colour ropes model)
two possible versions of string fusion mechanism were suggested.

The first version \cite{BP00EPJC} of the model assumes that
the colour fields are summing up only locally
in the area of overlaps of strings in the transverse plane.
So we will refer to this case as a {\it local} fusion
or {\it overlaps}. In this case one has
\beq
\av{n}{k}=\mu_0\frac{S_k}{\sig_0}\sqrt{k} \hs2
\av{p^2_t}{k}=\overline{p^2}\sqrt{k}
\label{local}
\eeq
Here $\av{n}{k}$ is the average multiplicity of charged particles
originated from the area $S_k$, where $k$ strings are overlapping,
and $\av{p^2_t}{k}$ is the same for their squared transverse momentum.
The $\mu_0$ and $\overline{p^2}$ are the average multiplicity
and squared transverse momentum of charged particles produced
from a decay of one single string,
and $\sig_0$ is its transverse area.

In the second version \cite{BMP02PRC} of the model one assumes that
the colour fields are summing up globally
- over total area of each cluster in the transverse plane -
into one average colour field.
This case corresponds to the summing of the source colour charges.
We will refer to this case as a {\it global} fusion
or {\it clusters}. In this case we have
\beq
\av{n}{cl}=\mu_0\frac{S_{cl}}{\sig_0}\sqrt{k_{cl}} \hs1
\av{p^2_t}{cl}=\overline{p^2}\sqrt{k_{cl}} \hs1
k_{cl}=\frac{ N^{str}_{cl} \sig_0 }{ S_{cl} }
\label{global}
\eeq
Here $\av{n}{cl}$ is the average multiplicity of charged particles
originated from the cluster of the area $S_{cl}$
and $\av{p^2_t}{cl}$ is the same for their squared transverse momentum.
The $N^{str}_{cl}$ is the number of strings forming the cluster.

\section{Cellular analog of SFM}

To simplify calculations in the case of real nucleus-nucleus collisions
a simple cellular model
originating from the string fusion model was proposed \cite{Vestn}.
In the framework of the cellular analog
along with the calculation simplifications
the asymptotics of correlation coefficients at large and small
string densities can be found analytically
in the idealized case with the homogeneous string distribution,
which enables to use these asymptotics later for the control of the
Monte-Carlo (MC) algorithms (see below).

Two versions of the cellular model as the original SFM
can be formulated - with local and global string fusion.
In this model we divide all transverse (impact
parameter) plane into sells of order of the transverse string size $\sig_0$.

In the version with {\it local} fusion
the assumption of the model is
that if the number of strings
belonging to the $ij$-th cell is $k_{ij}$, then they form
higher colour string, which emits in average $\mu _0\sqrt{k_{ij}}$ particles
with mean $p_t^2$ equal to $\overline{p^2}\sqrt{k_{ij}}$ (compare with (\ref{local})).
Note that zero "occupation numbers" $k_{ij}=0$ are also admitted.

In the version with {\it global} fusion at first
we define the neighbour cells as the cells with a common link.
Then we define the cluster as the set of neighbour cells
with non zero occupation numbers $k_{ij} \neq 0$.
After that we can apply the same formulae of the global fusion (\ref{global})
as in the original SFM,
where $N^{str}_{cl}$ is the number of strings in the cluster
and $S_{cl}/\sig_0$ is the number of cells in the cluster.

From event to event the number of strings $k_{ij}$ in the $ij$-th cell
will fluctuate around some average value - $\overline{k}_{ij}$.
Clear that in the case of real nuclear collisions
these average values $\overline{k}_{ij}$ will be different
for different cells. They will depend on the position ($\bfs_{ij}$)
of the $ij$-th cell in the impact parameter plane
(${\bf s}$ is two dimensional vector in the transverse plane).
In the case of nucleus-nucleus $AB$-collision
at some fixed value of impact parameter $\bfb$
one can find
this {\it average} local density of primary strings $\overline{k}_{ij}$
in the point $\bfs_{ij}$
using nuclear profile functions $T_A(\bfs_{ij}+\bfb/2)$ and $T_B(\bfs_{ij}-\bfb/2)$.

In MC approach
knowing the $\overline{k}_{ij}$ one can generate some configuration $C\equiv\{k_{ij}\}$.
To get the physical answer for one given event (configuration $C$)
we have to sum the contributions
from different cells in accordance with {\it local} or {\it global} algorithm
(see above),
which corresponds to the integration over ${\bf s}$ in transverse plane.
Then we have to sum over events (over different configurations $C$).
Note that as the event-by-event fluctuations of the impact parameter at a level of a few fermi
are inevitable in the experiment one has to include the impact
parameter $b$ into definition of configuration $C\equiv\{b,k_{ij}\}$.

\section{Long-range correlations}

The idea \cite{BP00EPJC,BP00PRL,ABP94PRL} to use
the study of long-range correlations in nuclear collisions
for observation of the colour string fusion phenomenon based
on the consideration that the quark-gluon string is an extended object
which fragmentation gives the contribution to wide rapidity range.
This can be an origin of the long-range correlations in rapidity space
between observables in two different and separated rapidity intervals.
Usually in an experiment
they choose these two separated rapidity intervals
in different hemispheres of the emission of secondary particles
one in the forward and another in the backward
in the center mass system.
So sometimes these long-range rapidity correlations are referred as
the forward-backward correlations (FBC).

In principle one can study three types of such long-range correlations:\\
$n$-$n$ - the correlation between multiplicities of charged particles
in these rapidity intervals,\\
$p_t$-$p_t$ - the correlation between transverse momenta
in these intervals and\\
$p_t$-$n$ - the correlation between the transverse momentum
in one rapidity interval and the multiplicity of charged particles
in another interval.

Usually to describe these correlations numerically
one studies the average value $\av{B}{F}$
of one dynamical variable $B$ in the backward rapidity window $\DB$,
as a function of another dynamical variable $F$
in the forward rapidity window $\DF$.
Here $\av{...}{F}$ denotes averaging over events
having a fixed value of the variable $F$ in the forward rapidity window.
The $\ave{...}$ denotes averaging over all events.
So we find the correlation function $\av{B}{F}=f(F)$.

It's naturally then to define the correlation coefficient
as the response of $\av{B}{F}$ on the variations of the variable
$F$ in the vicinity of its average value $\ave{F}$.
At that useful also to go to the relative variables,
i.e. to measure a deviation of $F$ from its average value $\ave{F}$
in units of $\ave{F}$, and the same for $B$.
So it's reasonable to define a correlation coefficient $b^{}_{B-F}$
for correlation between observables $B$ and $F$
in backward and forward rapidity windows in the following way:
\beq
b^{}_{B-F}\equiv\frac{\ave{F}}{\ave{B}} \left.\frac{d\av{B}{F}}{dF}\right|_{F=\ave{F}}
\label{bB-F}
\eeq

As the dynamical variables we use the multiplicity of charged particles ($n$),
produced in the given event in the given rapidity window,
and the {\it event(!)} mean value of their transverse momentum ($p_t$),
i.e. the sum of the transverse momentum magnitudes of all charged particles,
produced in the given event in the given rapidity window ($\Delta y$),
divided by the number of these particles ($n$):
\beq
p_t \equiv \frac{1}{n} \sum_{i=1}^{n} |{\bf p}_{ti}|,
\hs1 {\rm where} \hs1 y_i\in\Delta y;
\hs1 i=1,...,n.
\label{pt}
\eeq
So we can define three correlation coefficients: $b^{}_{n-n}$, $b^{}_{p_t-p_t}$ and $b^{}_{p_t-n}$,
for example:
\beq
b^{}_{p_t-p_t}\equiv\frac{\ave{\pF}}{\ave{\pB}}\left.\frac {d\av{\pB}{\pF}} {d\pF}
\right|_{\pF=\ave{\pF}}\ ,
\hs2
b^{}_{p_t-n}\equiv\frac{\ave{\F}}{\ave{\pB}}\left.\frac {d\av{\pB}{\F}} {d\F}
\right|_{\F=\ave{\F}}
\label{bpt-n}
\eeq
The $\B$, $\F$ are the multiplicities and $\pB$, $\pF$  are the
{\it event} (\ref{pt}) mean transverse momentum of the charged particles,
produced in the given event
correspondingly in the backward ($\DB$) and forward ($\DF$) rapidity windows.
In this paper we concentrate only on the $p_t$-$p_t$ - correlation.
Some other results can be found in \cite{dub04pro}.

\section{Analytical results and MC calculations}

In the framework of the cellular analog of SFM
in the idealized case with the homogeneous string distribution in the transverse plane
($\overline{k}_{ij}\equiv\eta=const$ for all $i$ and $j$)
the asymptotic of $p_t$-$p_t$ correlation coefficient
can be found analytically at large string density
$\rho^{str}=\eta/\sigma_0$, $\eta\gg1$,
when one cluster is forming:
\beq
b^{}_{p_t-p_t}=\frac{\mu_{0F}}{\mu_{0F}+16\gamma^2\,\sqrt{\eta}}
\hs1 {\rm and} \hs1
b^{}_{p^2_t-p^2_t}=\frac{\mu_{0F}}{\mu_{0F}+4\widetilde{\gamma}^2\,\sqrt{\eta}}
\label{pt-pt:cyl}
\eeq
Here $\mu_{0F}$ is the average number of charged particles produced
from a decay of one string in the forward rapidity window.
The $\gamma$ is the coefficient of proportionality between
the average transverse momentum $\overline{p}$
and the square root from the dispersion of $p$
for one string: $\sigma_{p}=\gamma\overline{p}$, where
$\sigma^2_{p}=\overline{p^2}-\overline{p}^2$.
The $\widetilde{\gamma}$ is the same for $p^2$:
$\sigma_{p^2}=\widetilde{\gamma}\overline{p^2}$.
In the last formula of (\ref{pt-pt:cyl}) as the dynamical variables
in the forward ($F$) and backward ($B$) rapidity windows instead of
$\pF$ and $\pB$ (\ref{pt}) we have used:
\beq
\ppF \equiv \frac{1}{\nF} \sum_{i=1}^{\nF} {\ppiF},
\hs1 {\rm and} \hs1
\ppB \equiv \frac{1}{\nB} \sum_{i=1}^{\nB} {\ppiB},
\label{pt2:FB}
\eeq
The asymptotics (\ref{pt-pt:cyl}) are similar to the ones
obtained in \cite{Vestn} for $n$-$n$ and $p_t$-$n$ correlations:
\beq
b^{}_{n-n}=b^{}_{p^2_t-n}=\frac{\mu_{0F}}{\mu_{0F}+4\sqrt{\eta}}
\hs1 {\rm and} \hs1
b^{}_{p_t-n}=\frac{1}{2}\ \frac{\mu_{0F}}{\mu_{0F}+4\sqrt{\eta}}\ ,
\label{pt-pt_cyl}
\eeq
notwithstanding the fact that
the derivation of (\ref{pt-pt:cyl}) is very different and much more complicated.
We see that for $p_t$-$p_t$ correlation in asymptotic the $\mu_0/\sqrt{\eta}$-scaling
obtained in \cite{Vestn} for $n$-$n$ and $p_t$-$n$ correlations
is also takes place. The correlation coefficients depends only on
the one combination of $\mu_0$ and $\eta$.



\begin{figure}[thb]
\centerline{\epsfig{figure=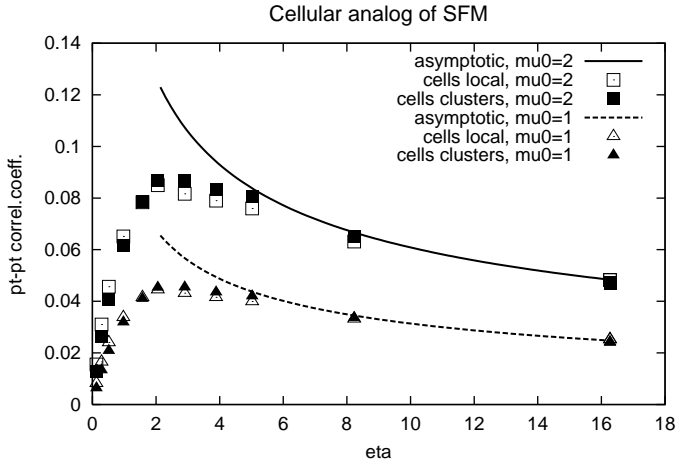,height=9cm,angle=-90}}
\caption{
The comparison of the asymptotic of $p_t$-$p_t$ correlation coefficient
$b^{}_{p_t-p_t}$ (\ref{pt-pt:cyl})
with the results of its MC calculations in the framework of the cellular analog of SFM
in the idealized case with the homogeneous string distribution in the transverse plane
at two values of $\mu_{0F}\equiv\mu_0$.
The $\eta$ is proportional to the string density $\eta=\sigma_0\rho^{str}$.
The total number of sells is $M=450$.
See the text for details.
}
\end{figure}
\begin{figure}[thb]
\centerline{\epsfig{figure=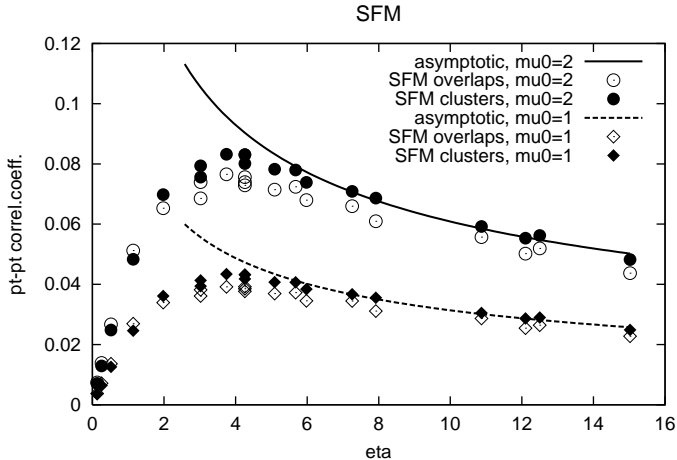,height=9cm,angle=-90}}
\caption{
The same as in Fig.1 but for
MC calculations in the framework of the original SFM
at the average number of strings $N^{str}=$1000, 4000 and 8000.
}
\end{figure}
\begin{figure}[thb]
\centerline{\epsfig{figure=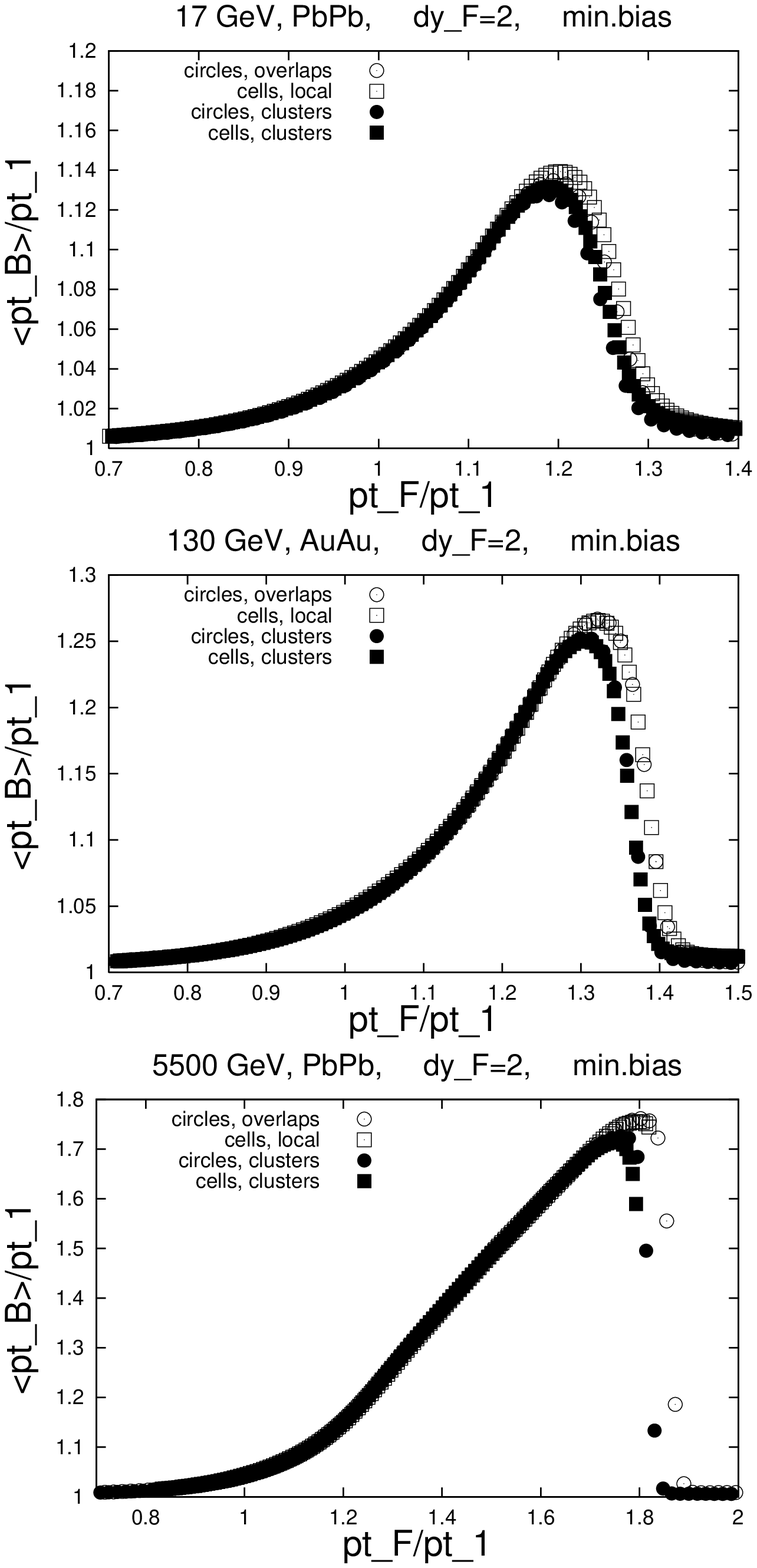,height=20cm}}
\caption{
The results of the MC calculations
of the $p_t$--$p_t$ correlation function $\av{\pB}{\pF}=f(\pF)$
for minimum bias nuclear collisions at SPS, RHIC and LHC energies.
See the text for details.
The results are presented in units of $pt_1$, where
$pt_1\equiv\overline{p}$ is the average transverse momentum
of particles produced from a decay of one string.
}
\end{figure}

In Fig.1 we compare the asymptotic (\ref{pt-pt:cyl}) of $b^{}_{p_t-p_t}$
with the results of its MC calculations in the framework of the cellular analog of SFM
described above. We see that at $\eta>5$ the results
for both {\it local} and {\it global} versions
of the model practically coincide with the asymptotic.
The MC calculations in Fig.1 fulfilled for the total number of cells $M=450$,
but the simulations with the small number of cells $M=45$
give practically the same results.
So for $p_t$-$p_t$ correlation one also has "$M$-scaling",
which was found earlier in \cite{Vestn} for $p_t$-$n$ and $n$-$n$ correlations.
This means that the $p_t$-$p_t$ correlation coefficient depends only
on the string density $\rho^{str}=\eta/\sigma_0$ but not on the total number of
strings $N^{str}=\eta\,M$.

In Fig.2 we compare the same asymptotic (\ref{pt-pt:cyl})
with the results of MC calculations in the framework of the original SFM
(with the taking into account the real geometry of merging strings,
which takes much more efforts).
Again we see that at $\eta>5$ the MC results
calculated for {\it local (overlaps)} and {\it global (clusters)}
versions of the original SFM practically coincide with the asymptotic (\ref{pt-pt:cyl}).
In Fig.2 the points correspond to the MC calculations fulfilled
at the different number of strings $N^{str}=$1000, 4000 and 8000.
We see that in this case the results also do not depend on the total
number of strings but depend only on the string density $\rho^{str}=\eta/\sigma_0$.

We would like to emphasize
that this takes place only due to the fact that as observables $\pF$ and $\pB$
we choose the event mean values of transverse momenta of the particles emitted
in the rapidity intervals (see (\ref{pt}) and (\ref{pt2:FB})).
In paper \cite{EPJC04} it was shown that if one takes as observables
the transverse momenta of single particles produced in these two rapidity windows
(two particle correlation)
then such defined $p_t$--$p_t$ correlation coefficient
is inversely proportional to the
number of strings $N^{str}$ and rapidly decreases to zero with the increase
of the total number of strings.

Comparing the Fig.1 and Fig.2 we see also that the MC calculations
in the framework of the original SFM and in the framework of its cellular analog
give very similar results at all values of string density.

\section{The $p_t$-$p_t$ correlation in minimum bias AA-collisions}
In the case with the realistic nucleon distribution density of colliding nuclei
the results of the MC calculations
of the $p_t$--$p_t$ correlation function
(see section 3): $\av{\pB}{\pF}=f(\pF)$
for minimum bias nuclear collisions at SPS, RHIC and LHC energies
are presented in Fig.3.
Certainly, in the case of minimum bias MC calculations one has to include the impact
parameter $b$ into definition of configuration $C$ (see section 2).
In these figures ($\circ$) and ($\bullet$) denote the results of calculations
in the framework of the original SFM
for its {\it local (overlaps)} and {\it global (clusters)}
versions correspondingly.
The ($\Box$) and (\rule[0.5mm]{2mm}{2mm}$\,$) denote the results of calculations
in the framework of the cellular analog of SFM
for its {\it local} and {\it global (clusters)}
versions. All presented results are
for the forward rapidity window of 2 unit length ($\DF=2$).

First of all we see in Fig.3 that the results of MC calculations for all four versions
are practically coincide with each other.
Only the results for both versions with local fusion are slightly shifted
to the higher values of $p_t$ in comparison with the ones for clusters,
which can be easily explained by simple general arguments.

We see also that in the case of minimum bias nuclear collisions
the $p_t$--$p_t$ correlation functions $\av{\pB}{\pF}$
exhibit strongly nonlinear behavior.
Note that the correlation functions calculated at fixed
value of impact parameter (or with its small fluctuations within a certain centrality class)
are practically linear and the strength of the correlation can be described
by one number - the correlation coefficient (see (\ref{bpt-n})).
It's not true for minimum bias collisions and the correlation can't be described
by one number in this case - the derivative of the correlation function is
positive at small values of $\pF$ and it becomes negative at large values of $\pF$.

This can be explained in the following way. The set of minimum bias collisions
includes both central and peripheral collisions.
In our model with string fusion due to higher density of strings
one has higher value of $\pB$ and $\pF$  for central collisions,
than for peripheral ones.
This is the reason for the positive $p_t$--$p_t$ correlation.
To understand the reason for the negative $p_t$--$p_t$ correlation at large values of $\pF$
one has to remember that
as observables $\pF$ and $\pB$ (\ref{pt}) (\ref{pt2:FB})
we choose
the event mean values of transverse momenta of the particles emitted
in the rapidity intervals.
Only in this case the $p_t$--$p_t$ correlation is not equal to zero
in real heavy ion collisions (see the discursion in section 4).
It's  clear from (\ref{pt}) (\ref{pt2:FB}) that although the event mean values
of $\pF$ and $\pB$ are higher for central collisions, their dispersions is higher
for peripheral ones,
as the number of charged particles $\nF$ and $\nB$ produced in
these rapidity intervals are smaller for peripheral collisions.
This means that if in some event one observes a very high (or a very low) value of $\pF$
in the forward rapidity window (than the typical value $\ave{\pF}$),
then it's more probably that the peripheral
collision has taken place and then one finds the $\pB$ in the backward
rapidity window at a low level - typical for peripheral collisions.
So with the increase of $\pF$ the function $\av{\pB}{\pF}$ starts and finishes
at a low (typical for peripheral collisions) value of $\pB$ obtaining
in the middle more higher (typical for central collisions) value.

Using these arguments and taking into account the increase of string density,
typical values of $\pF$ and $\pB$, and the multiplicities $\nF$ and $\nB$,
with the increase of the initial energy, one enables, also to explain
the change of the form of the $p_t$--$p_t$ minimum bias correlation function in Fig.3
from SPS to RHIC and LHC energies.

\section{Conclusion}
The theoretical description of the correlations between observables
in two separated rapidity intervals in relativistic nuclear collisions
is presented.
The calculations are fulfilled in the framework of the model
taking into account the possibility of quark-gluon string fusion
and also in the framework of the suggested simple cellular analog of this
model. For both models two possible mechanisms of a string
merging are considered: the local fusion ("overlaps") and
the global fusion (with forming  of colour "clusters").

It is shown, that for the $p_t$--$p_t$ correlation the event-by-event correlation
between event mean values of transverse momenta of the particles emitted
in two different rapidity intervals does not decrease to zero
with the increase of the number of strings
in contrast with the correlation between the transverse momenta
of single particles produced in these two rapidity windows
which was studied earlier \cite{EPJC04}.

In the idealized case with the homogeneous string distribution in the transverse plane
the asymptotic of $p_t$--$p_t$ correlation coefficient
analytically calculated in the framework of global string fusion mechanisms,
when at large string density one cluster is forming.
The asymptotic is compared with
the results of the MC calculations in this idealized case
both in the framework of the original SFM and in the framework of its cellular analog,
which enables to control the MC algorithms.

In the case with the realistic nucleon distribution density of colliding nuclei
the results of the MC calculations of the $p_t$--$p_t$ correlation function
for minimum bias nuclear collisions at SPS, RHIC and LHC energies
are presented and analysed.

\subsection*{Acknowledgments}

The authors would like to thank M.A.~Braun
and G.A.~Feofilov for numerous valuable
encouraging discussions.
The work has been supported
by the grant 05-02-17399-a of the Russian Foundation for Fundamental Research
and by the grant A04-2.9-147 of the Education and Science Ministry of Russia.


%

\end{document}